
\magnification=1200
\baselineskip=16pt

\pageno=0
\rightline{ILL-(TH)-92-6}
\rightline{CERN-TH.6515/92}

\centerline{ \bf THE UNIVERSALITY CLASS OF MONOPOLE CONDENSATION}
\centerline{\bf IN NON-COMPACT, QUENCHED LATTICE QED}
\vskip2truecm

\centerline{Aleksandar KOCI\' C\footnote{$\,^{ }$}{ILL-(TH)-92-6} }
\centerline{\it Theory Division, CERN, CH-1211 Geneva 23,
Switzerland}

\vskip1truecm

\centerline{John B. KOGUT\footnote{$\,^{ }$}{CERN-TH.6515/92}}
\centerline{\it Department of Physics, University of
Illinois at Urbana-Champaign}
\centerline{ \it 1110 West Green Street, Urbana,
IL 61801-3080 USA}
\vskip1truecm

\centerline{Simon J. HANDS}
\centerline{\it Department of Physics and Astronomy,
University of Glasgow, Glasgow G12 8QQ,  UK}
\centerline{\it and}
\centerline{\it Theory Division, CERN, CH-1211 Geneva 23, Switzerland}

\vskip2truecm

\centerline{\bf Abstract}

{\narrower
Finite size scaling studies of monopole condensation in noncompact
quenched lattice $QED$ indicate an authentic second order phase transition
lying in the universality class of four dimensional percolation.  Since the
upper critical dimension of percolation is six, the measured critical
indices are far from mean-field values.  We propose a simple set of ratios
as the exact critical indices for this transition.  The implication of
these results for critical points in Abelian gauge theories are discussed.
\footnote{$\,^{ }$}{June 1992}
\smallskip}
\vfill\eject

Monopole condensation was identified long ago as the physical mechanism
driving the confinement transition in $U(1)$ lattice gauge theory [1].  This
model illustrates the dual Meissner effect which is presumed to occur in
non-Abelian theories as discussed by t'Hooft [2] and Mandelstam [3].  The
complexity of these models, however, has led to slow progress in sharpening
our understanding of the condensation mechanism and its physical
implications.  It is, therefore, interesting to consider a particularly
simple model of monopole condensation in four dimensions, where exacting
work can be done.  In particular, we shall consider monopole condensation
in non-compact quenched lattice $QED$ and argue that it is in the same
universality class as four dimensional percolation.  The simplicity of the
model will allow us to do accurate numerical studies, which will determine
the monopole susceptibility critical index $\gamma$, the monopole percolation
(``magnetic'' critical index $\beta$,
and the correlation length critical index $\nu$
to a few percent.  We shall see that the critical indices coincide with
those of four dimensional percolation.  These results are interesting
because they indicate that monopole condensation in non-compact quenched $QED$
is an authentic second order phase transition even though the local field
theory in which it is embedded is just a free field.  In addition, since
the upper critical dimensionality of percolation is six, the critical
indices associated with the phase transition are far from mean-field
values.
        We begin by discussing our numerical determinations of the critical
indices $\gamma, \beta$ and $\nu$. The lattice action we simulated is

$$
S_{gauge}={1\over 2}\sum_{n\mu\nu}
\bigl( \theta_\mu(n) +\theta_\nu(n+\mu)
-\theta_\mu(n+\nu)-\theta_\nu(n)\bigr)^2
\equiv {1\over 2} \sum_{n\mu\nu} \Theta_{\mu\nu}^2(n),
\eqno(1)
$$
where the gauge fields $\theta_\mu(n)$
are oriented, real variables in the range $(- \infty,
+ \infty)$
defined on lattice links.  Although Eq.(1) is just a free field, we can
define a magnetic charge on the lattice as was already done in
compact lattice $QED_4$.  Introduce an electric charge $e$ and define an
integer-valued Dirac string by,

$$
e\Theta_{\mu\nu}(n)=e\bar\Theta_{\mu\nu}(n) + 2\pi S_{\mu\nu},
\eqno(2)
$$
where the integer $S_{\mu\nu}$
determines the strength of the string threading the
plaquette and $e\bar\Theta_{\mu\nu}$ is defined to lie in the interval
$(-\pi, +\pi]$.
The integer-valued monopole current $m_\mu(\tilde n)$, defined on links of the
dual lattice, is then

$$
m_\mu(\tilde n)= {1\over 2} \epsilon_{\mu\nu\kappa\lambda}
\Delta_\nu^+ S_{\kappa\lambda}(n+\hat\mu)
\eqno(3)
$$
where $\Delta_\nu^+$ is the forward lattice difference operator, and
$m_\mu$ is the
oriented sum of the $S_{\mu\nu}$ around the faces of an elementary cube.  This
definition, which is gauge-invariant, implies the conservation law
$\Delta_\mu^- m_\mu(\tilde n)=0$
which means that monopole world lines form closed loops.
A fuller discussion of these variables can be found in ref.[5] where useful
contrasts are made with the same constructions in compact lattice $QED$.

As emphasized originally in ref.[5], the constructions and concepts of
percolation [6-8] are useful in quantifying the meaning of monopole
 condensation.
  Introduce the idea of a connected cluster
of monopoles on the dual lattice:  one counts the number of dual sites
joined into clusters by monopole line elements.  The oriented (vector-like)
nature of the monopole elements is ignored.  The problem of identifying and
counting clusters is now exactly the same as occurs in bond percolation
problems [6-8].  In the simplest models of bond percolation the entire
problem is one of counting.  One assumes that bonds are occupied randomly
with probability $\rho$.  At some critical concentration $\rho_c$ (called the
percolation threshold), the largest connected cluster becomes infinite in
extent, occupying a macroscopic fraction of the dual lattice, and
signalling a phase transition.  A natural order parameter for the phase
transition is $M = n_{max}/n_{tot}$ where $n_{max}$
is the number of sites in the
largest cluster and ntot is the total number of connected sites [8].  Its
associated susceptibility reads,

$$
\chi={
{<\sum_{n_{min}}^{n_{max}} g_n n^2 - n_{max}^2>}\over {n_{tot}}
}
\eqno(4)
$$
where $n$ labels the size of a cluster occuring $g_n$ times on the dual
lattice.  In general $n_{min} = 2$, but for monopoles $n_{min} = 4$
because of the conservation law.

The critical indices for monopole condensation are then defined as in bond
percolation[8].  For a lattice of infinite extent, $M$ should be nonzero for a
strong coupling e and vanish identically for weak coupling.  At some
critical point, $M$ should turn on non-analytically with a "magnetic"
exponent $\beta$,

$$
M\sim\biggl({1\over{e^2_c}}-{1\over{e^2}}\biggr)^\beta,
\,\,\,\,\,\,
e\geq e_c
\eqno(5)
$$
where we have written $1/e^2$ (rather than $e$ itself), following the standard
conventions of strong coupling lattice gauge theory [1]. The susceptibility
should diverge at $e_c$ with a susceptibility index $\gamma$

$$
\chi\sim\biggl({1\over{e^2_c}}-{1\over{e^2}}\biggr)^{-\gamma}
\eqno(6)
$$
and the linear size $\xi$ of the largest cluster should also diverge:

$$
\xi\sim\biggl({1\over{e^2_c}}-{1\over{e^2}}\biggr)^{-\nu},\,\,\,\,\,\,
e\geq e_c
\eqno(7)
$$
where $\nu$ is the correlation length exponent.

In order to measure $\gamma,\beta$ and $\nu$,
we did three types of computer experiment.
First we measured $\chi$ and $M$ as a function of coupling $1/e^2$
on lattices of
volume $L^4$, with $L$ ranging from 10 to 20.  Since $S_{gauge}$
is a quadratic form,
independent gauge field configurations could be generated by FFT methods
avoiding critical slowing down entirely [9].  According to standard
finite-size scaling arguments [10], the peak of the susceptibility should
grow with lattice size $L$ as

$$
\chi_{max}(L)\sim L^{\gamma/\nu}.
\eqno(8)
$$
In addition, the value of the order parameter $M$ should vanish as

$$
M(L)\sim L^{-\beta/\nu},\,\,\,\,\,\, e=e_c
\eqno(9)
$$
at the critical point.  And finally, Eq.(5) yields an estimate of the
magnetic exponent $\beta$ as long as we can study a range of coupling
$1/e^2$ where
the functional dependence of Eq.(5) is not distorted by finite size
effects.  Consider the susceptibility first.  Data of $\chi$ versus $1/e^2$ for
$L$ values 10, 12, 16, 18 and 20 are given in Table 1.  Note that the peak
occurs at $1/e_c^2 = 0.244$ independent of $L$.
In Fig. 1 we plot the logarithms
of the peaks, $\ln \chi_{max}$ versus $\ln L$,
and find an excellent straight-line fit with the slope,

$$
\gamma/\nu =2.24(2)
\eqno(10)
$$
The order parameter $M$ was measured in parallel with $\chi$,
and the results are
recorded in Table 2.  A plot of $\ln M$ as a function of $\ln L$ at
$1/e_c^2 = 0.244$ is shown in
Fig. 2.  Again a straight line fit compatible with finite-size scaling
emerges and the slope is determined to be

$$
\beta/\nu = 0.88(2)
\eqno(11)
$$

Finally we attempted a direct measurement of Eq.(5) using the $16^4, 18^4$ and
$20^4$ lattices.  The measurements of $M$ indicate that its value
at $1/e^2 = 0.242$
decreases significantly
as $L$ ranges from 16 to 20, and are not useful here.  However, the couplings
$1/e^2 =0.240, 0.238, 0.236$ yield stable $M$ values.  We plot $\ln M$ against
$\ln (1/e_c^2 - 1/e^2)$
in Fig.3, and find that a linear fit is acceptable for the
three points $1/e^2 = 0.236, 0.238, 0.240$. The fit gives an estimate for the
slope, the magnetic exponent $\beta$,

$$
\beta=0.58(2)
\eqno(12)
$$

Clearly the measurements of $\gamma/\nu$ and
$\beta/\nu$ in Eqs.(10) and (11) are our most
precise.  It is interesting to test whether the critical indices of the
monopole condensation transition satisfy hyperscaling (which is expected of
any model with a single divergent correlation length, which controls the
system's non-analyticities at the critical point), and deduce other critical
exponents of the transition.  According to hyperscaling,

$$
\gamma/\nu = 2-\eta
\eqno(13a)
$$
$$
\beta/\nu =(d-2+\eta)/2,
\eqno(13b)
$$
where $d = 4$ here.  From Eq.(13.a) the critical index $\eta = -0.24$
and from Eq.(13.b) $\eta = -0.24$, as well.
The agreement with hyperscaling is perfect!  The third
hyperscaling relation $2\beta/\nu + \gamma/\nu = d$,
becomes $1.76 + 2.24 = 4.00$ and works perfectly also, while the fourth
hyperscaling relation, $2\beta\delta-\gamma=d\nu$,
gives the critical index
$\delta = 3.55(2)$.
Finally, the fifth and last hyperscaling relation, $d\nu = 2 -\alpha$,
requires additional input
for the specific heat index $\alpha$.
Now use Eq.(12), our determination of the magnetic
exponent.  For example, combining Eqs.(10) and (12) with the hyperscaling
relations gives all the critical indices of the transition,

$$
\gamma = 1.48(3), \,\,\,\,\,\,
\nu = 0.66(3), \,\,\,\,\,\,
\eta = -0 .24(2),
$$
$$
\alpha = -0.64(3), \,\,\,\,\,\,
\beta = 0.58(2),\,\,\,\,\,\,
\delta = 3.55(3)
\eqno(14)
$$
We complete the discussion of our numerical results with two observations.
First, we should classify this critical point, if at all possible.
Naturally, we suspect that it is simply related to a four-dimensional
percolation problem, although we have not proved this analytically, because
of the vector character of the monopole problem and the conservation law
$(\Delta^-_\mu m_\mu(\tilde n))$.
In fact, the results of Eq.(14) are in
excellent agreement with the critical exponents of four-dimensional
percolation, as estimated both by numerical simulation [6] and from series
expansions [7], so we suggest that these two transitions lie in the same
universality class.  This is particularly intriguing since the indices of
Eq.(14) are far from mean field indices.  In fact, the upper critical
dimensionality of percolation is 6, where mean field considerations become
exact and $\gamma = \beta= 1,
\nu = 1/2, \eta = 0, \alpha = -1$ and $\delta = 2$ [11].
Second, since the results of Eq.(14)
are so close to ratios of small integers, we conjecture that the exact
critical indices of this universality class are,

$$
\gamma = 3/2, \,\,\,\,\,\,
\nu = 2/3,  \,\,\,\,\,\,
\eta = -1/4,
$$
$$
\alpha = -2/3,\,\,\,\,\,\,
\beta = 7/12,\,\,\,\,\,\,
\delta = 25/7
\eqno(15)
$$

A comment about the negative value of $\eta$ is in order.  The hyperscaling
relations (Eq.13) are derived on the assumption that the physics of the
critical region can be described by a local scalar field theory.  In
general for a field of dimension $d_\phi$ they read:

$$
\gamma/\nu = d-2d_\phi, \,\,\,\,\,\,
\beta/\nu=d_\phi,
\eqno(16)
$$
which for $d_\phi = (d - 2 + \eta)/2$
reproduces Eq.(13).  If $\eta$ is negative, we see
that $d_\phi$ is smaller  than the canonical value $d/2 - 1$, which leads to
an infrared behaviour that is
inconsistent with unitarity.  The numerical success of
the hyperscaling relations, the conjectured exact fractional form of the
critical indices Eq.(15), and the field theoretic description of
the critical point remain to be integrated into a single comprehensive
theory of percolation.

The fact that quenched non-compact $QED$ monopoles condense with the same
exponents as four-dimensional percolation is perhaps not too surprising.
In ref.[5] it was shown that the concentration of monopole world lines
varies smoothly as $1/e^2$ is made to vary across the critical region,
i.e. there
exists a smooth function $\rho(1/e^2)$ for the probability of bond occupation,
and so the only source of non-analytic behavior can be the percolation
threshold itself.  Dynamical considerations are entirely subsumed by
geometrical ones.  This is in marked contrast to the compact $U(1)$ model, in
which the monopole density falls sharply across the deconfining
transition [4].  It would be interesting to repeat these measurements in
non-compact lattice $QED$ including the effects of dynamical fermions.  We
have argued elsewhere [12] that in this case, because of the compact nature of
the $U(1)$ connection required to couple fermions to the model, the monopoles
might have a direct dynamical influence on the physics of chiral symmetry
breaking - there is no reason {\it a priori}  to expect that monopole
condensation in this case lies in the same universality class.  Preliminary
results with $N_f = 2$ suggest that
the two cases are difficult to distinguish [13].
Work continues on this interesting  problem.
\vskip1truecm

JBK is supported in part by grant NSF-PHY87-01775.  SJH is supported in
part by an S.E.R.C. Advanced Fellowship.
\vfill\eject
\noindent
{\bf References}
\vskip5truemm

1.      T. Banks, R. Myerson and J.B. Kogut, Nucl. Phys. {\bf B129}, 493
(1977).

\noindent
2.      G. 't Hooft, in "High Energy Physics", ed. A. Zichichi, Palermo
        (Editrice Compositori, Bologna, 1976).

\noindent
3.      S. Mandelstam, Phys. Rep. {\bf C23}, 245 (1976).

\noindent
4.      T.A. DeGrand and D. Toussaint, Phys. Rev. {\bf D22}, 2478 (1980).

\noindent
5.      S. Hands and R. Wensley, Phys. Rev. Lett. {\bf 63}, 2169 (1989).

\noindent
6.      S. Kirkpatrick, Phys. Rev. Lett. {\bf 36}, 69 (1976).

\noindent
7.      D.S. Gaunt and H. Ruskin, J. Phys. {\bf A11}, 1369 (1978).

\noindent
8.      D. Stauffer, Phys. Rep. {\bf 54}, 1 (1979).

\noindent
9.      E. Dagotto, A. Koci\' c
and J.B. Kogut, Nucl. Phys. {\bf B317}, 253 (1989).

\noindent
10.     M.N. Barber, in "Phase Transitions", ed. M.S. Green, London
        (Academic Press, London 1983).

\noindent
11.     G. Toulouse, Laboratoire de Physique des Solides, Orsay, Report No.
110, 1973       (unpublished).

\noindent
12.     S.J. Hands, J.B. Kogut and A. Koci\' c,
Nucl. Phys. {\bf B357}, 467 (1991).

\noindent
13.     S.J. Hands, J.B. Kogut, R. Renken, A. Koci\' c, D.K. Sinclair and K.C.
Wang,   Phys. Lett. {\bf B261}, 294 (1991).

\eject
\centerline{\bf Table 1}

\noindent
Monopole susceptibility $\chi$ as a function of coupling $1/e^2$ for lattices
$10^4, 12^4, 16^4, 18^4$ and $20^4$.
\vskip 1 truecm
$$\vbox{\settabs\+\quad.244\quad&\quad52.02(97)\quad&\quad70.85(93)\quad&
\quad146.2(3.5)\quad&\quad192.9(3.4)\quad&\quad248.8(7.4)\quad&\cr
\+\hfill$1/e^2$\hfill&\hfill$10^4$\hfill&\hfill$12^4$\hfill&\hfill$16^4$\hfill&
\hfill$18^4$\hfill&\hfill$20^4$\hfill&\cr
\bigskip
\+\hfill.254\hfill&\hfill38.66(38)\hfill&\hfill50.27(45)\hfill&\hfill67.35(60)
\hfill&\hfill72.80(91)\hfill&\hfill75.2(1.5)\hfill&\cr\smallskip
\+\hfill.252\hfill&\hfill41.89(45)\hfill&\hfill56.64(57)\hfill&\hfill75.58(78)
\hfill&\hfill88.3(1.2)\hfill&\hfill94.8(2.2)\hfill&\cr\smallskip
\+\hfill.250\hfill&\hfill45.93(57)\hfill&\hfill63.44(72)\hfill&\hfill95.6(1.0)
\hfill&\hfill109.1(1.7)\hfill&\hfill117.5(2.9)\hfill&\cr\smallskip
\+\hfill.248\hfill&\hfill49.16(72)\hfill&\quad70.85(93)\quad&\hfill116.4(1.6)
\hfill&\hfill140.7(2.9)\hfill&\hfill153.1(4.8)\hfill&\cr\smallskip
\+\hfill.246\hfill&\hfill51.37(83)\hfill&\hfill75.9(1.1)\hfill&\hfill136.9(2.5)
\hfill&\hfill172.9(2.5)\hfill&\hfill211.4(4.3)\hfill&\cr\smallskip
\+\quad.244\quad&\quad52.02(97)\quad&\hfill77.2(1.4)\hfill&\quad146.2(3.5)
\quad&\quad192.9(3.4)\quad&\quad248.8(7.4)\quad&\cr\smallskip
\+\hfill.242\hfill&\hfill51.9(1.1)\hfill&\hfill73.8(1.6)\hfill&\hfill134.9(4.4)
\hfill&\hfill167.4(5.0)\hfill&\hfill215.8(10.3)\hfill&\cr\smallskip
\+\hfill.240\hfill&\hfill48.2(1.1)\hfill&\hfill66.2(1.8)\hfill&\hfill106.4(4.5)
\hfill&\hfill114.5(9.6)\hfill&\hfill105.7(12.8)\hfill&\cr\smallskip
\+\hfill.238\hfill&\hfill43.1(1.3)\hfill&\hfill56.9(1.9)\hfill&\hfill63.8(3.2)
\hfill&\hfill64.6(5.0)\hfill&\hfill62.8(6.1)\hfill&\cr\smallskip
\+\hfill.236\hfill&\hfill36.3(1.2)\hfill&\hfill42.1(1.6)\hfill&\hfill38.7(2.1)
\hfill&\hfill35.7(1.4)\hfill&\hfill37.1(3.1)\hfill&\cr}$$
\vfill\eject
\centerline{\bf Table 2}

\noindent
Same as Table 1, but for the order parameter $M = n_{max}/n_{tot}$.
\vskip 1 truecm
$$\vbox{\settabs\+\quad.244\quad&\quad.2172(33)\quad&\quad.1882(28)\quad&
\quad.1410(29)\quad&\quad.1283(23)\quad&\quad.1165(29)\quad&\cr
\+\hfill$1/e^2$\hfill&\hfill$10^4$\hfill&\hfill$12^4$\hfill&\hfill$16^4$\hfill&
\hfill$18^4$\hfill&\hfill$20^4$\hfill&\cr
\bigskip
\+\hfill.254\hfill&\hfill.0952(15)\hfill&\hfill.0652(11)\hfill&\hfill.0311(60)
\hfill&\hfill.0225(6)\hfill&\hfill.0165(7)\hfill&\cr\smallskip
\+\hfill.252\hfill&\hfill.1096(19)\hfill&\hfill.0776(13)\hfill&\hfill.0392(8)
\hfill&\hfill.0282(9)\hfill&\hfill.0196(9)\hfill&\cr\smallskip
\+\hfill.250\hfill&\hfill.1282(22)\hfill&\hfill.0942(16)\hfill&\hfill.0507(11)
\hfill&\hfill.0368(13)\hfill&\hfill.0282(17)\hfill&\cr\smallskip
\+\hfill.248\hfill&\hfill.1519(26)\hfill&\hfill.1162(19)\hfill&\hfill.0684(16)
\hfill&\hfill.0518(19)\hfill&\hfill.0390(27)\hfill&\cr\smallskip
\+\hfill.246\hfill&\hfill.1809(29)\hfill&\hfill.1468(23)\hfill&\hfill.0968(21)
\hfill&\hfill.0816(16)\hfill&\hfill.0688(20)\hfill&\cr\smallskip
\+\quad.244\quad&\quad.2172(33)\quad&\quad.1882(28)\quad&\quad.1410(29)
\quad&\quad.1283(23)\quad&\quad.1165(29)\quad&\cr\smallskip
\+\hfill.242\hfill&\hfill.2559(37)\hfill&\hfill.2374(31)\hfill&\hfill.2052(33)
\hfill&\hfill.1993(27)\hfill&\hfill.1950(34)\hfill&\cr\smallskip
\+\hfill.240\hfill&\hfill.3033(38)\hfill&\hfill.2949(33)\hfill&\hfill.2775(34)
\hfill&\hfill.2722(55)\hfill&\hfill.2792(67)\hfill&\cr\smallskip
\+\hfill.238\hfill&\hfill.3557(39)\hfill&\hfill.3530(32)\hfill&\hfill.3518(28)
\hfill&\hfill.3505(38)\hfill&\hfill.3497(52)\hfill&\cr\smallskip
\+\hfill.236\hfill&\hfill.4099(37)\hfill&\hfill.4147(28)\hfill&\hfill.4154(22)
\hfill&\hfill.4162(29)\hfill&\hfill.4135(42)\hfill&\cr}$$
\vfill\eject
\centerline{\bf Figure captions}

\noindent
1. Finite-size scaling for the peaks of the monopole susceptibility
$\ln \chi_{max}$ as a function of $\ln L$.

\noindent
2. Same as Fig.1, but for the order parameter $M$ at $e_c$.

\noindent
3. $\ln M$ against $\ln (1/e_c^2 - 1/e^2)$
for $1/e^2 = 0.242, 0.240, 0.238$ and $0.236$.  The
size of the symbols for $16^4, 18^4$ and $20^4$
lattices include the statistical
error bars except in the case $1/e^2 = 0.242$, which is not used in the fit
because of the finite-size effects.
\vfill\end